\documentclass[aps,twocolumn,superscriptaddress]{revtex4}
\usepackage{amssymb}
\usepackage{amsmath}
\usepackage{graphicx}
\usepackage{subfigure}
\usepackage{color}
\begin{document}

\title {Rectification Effect on Solitary Waves in the Symmetric Y-shaped Granular Chain}

\author{Xingyi Liu} \author{Tengfei Jiao} \author{Jiaye Su}
\affiliation{Department of Applied Physics, Nanjing University of Science and Technology, Nanjing 210094, China}

\author{Weizhong Chen}
\affiliation{Key Laboratory of Modern Acoustics of Ministry of Education, Institute of Acoustics, Nanjing University, Nanjing 210093, China}

\author{Qicheng Sun}
\affiliation{State Key Laboratory of Hydroscience and Engineering, Tsinghua University, Beijing 100084,China}

\author{Decai Huang}
\email[Corresponding author:] {hdc@njust.edu.cn}

\affiliation{Department of Applied Physics, Nanjing University of Science and Technology, Nanjing 210094, China}
\affiliation{Key Laboratory of Modern Acoustics of Ministry of Education, Institute of Acoustics, Nanjing University, Nanjing 210093, China}
\affiliation{State Key Laboratory of Hydroscience and Engineering, Tsinghua University, Beijing 100084,China}

\date{\today}

\begin{abstract}

The rectification effect on the propagation of solitary waves in the symmetric Y-shaped granular chain is numerically investigated in this Letter. A heterojunction with mass mismatch occurs at the position of Y-junction by adjusting the branch angle. And the heavy-light heterojunction is more favorable for the solitary wave passing. The energy rectification efficiency can be improved by adjusting the branch angle and the direction of incident solitary wave. The results have particularly practical significance for the potential design of acoustic diode devices.

\end{abstract}

\maketitle

The device that has rectification effect on the energy flux such as the electrical, magnetic and thermal diodes has been a hot topic in both scientific and technical communities \cite{Tongay2012PRL2P011002, Zietek2016APL109P072406, Lepri2003PR377P1}. These pioneering works, especially for the electrical diodes, have brought dramatic revolutions in various fields. In the past decades, the acoustic diode has been attracting extensive investigations, which provides huge potential applications such as the acoustic silencers \cite{Zhu2011PRL106P014301}, mechanical dampers \cite{Sen2011APL99P063501}, and energy containers \cite{Sen2015APL107P244105}. The granular acoustic diode (GAD) has been identified in one-dimensional granular chain by several experiments and simulations \cite{Sen2008PR462P21, Vergara2005PRL95P108002, Daraio2012PRE85P036602, Daraio2014JMPS73P103}. For the potential design and application of GAD, it is crucial to construct a basic block of heterojunction and to learn about the rectification efficiency.

\begin{figure}[htbp]\centering
  \includegraphics[width=9.0cm, trim=0.0cm 7.0cm 0.0cm 7.0cm, clip,
   angle=0]{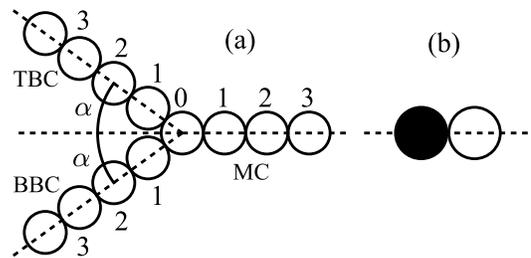}
  \caption { Sketch of (a) simulation system and (b) Quasi-particle model at the branch angle of $\alpha=30^{0}$.}
  \label{fig:Figure1Model}
\end{figure}

In this Letter, we numerically study the rectification effect on the solitary wave (SW) in a symmetric Y-shaped granular chain (YGC). The simulation results demonstrate the possibility for establishing a simple model of GAD. The YGC is consisted of one main chain (MC) and two branch chains (BCs) of the top (TBC) and bottom branch chain (BBC), shown in Fig. \ref{fig:Figure1Model}(a). In such a system, the mass mismatch at the Y-shaped junction results in a heterojunction, which allows us to tune the transmission and reflection.

In the simulations, the branch angle ($\alpha$) between the TBC and MC is equally set as that of BBC with MC. Those of three chains share one of the same grain (marked by grain 0) at the interface of Y-shaped junction and each of the MC and two BCs individually has 50 grains (marked by 1,2, $\cdots$, N ) which are spherical and placed in three lines separately. And at the beginning of simulations, the grains are arranged in barely touching each other. The method of molecular dynamics is applied and the dynamics of each grain is governed by Newton's equations as used in our previous studies \cite{HuangPRE2006, HuangEPJE2013}. In a simulation time step, the position and velocity of grains are updated in turn. Only the translational motion is considered in the simulations. The normal interaction of two contact grains is determined by Hertz's law in the absence of dissipation \cite{Kuwabara1987, Schafer1996}.

\begin{align}
F_{i,i+1}=\left
\{\begin{array}{l}
kz^{3/2}_{i, i+1},~~z_{i, i+1}\geq0 \\
~~~0,~~~~~~ z_{i, i+1}<0  \\
\end{array}\right.
\label{eq:Force}
\end{align}%

Herein, the overlap of two touching grains is denoted as $z_{i, i+1}=d-(x_{i+1}-x_{i})$, wherein the grain diameter is $d=5.0~{\rm mm}$ , and the absolute positions of grain $i$ and grain $i+1$ at the time of $t$ are $x_i$ and $x_{i+1}$ , respectively. Also the elastic coefficient $k$ is calculated as the following: $k=Yd^{\frac{1}{2}}/3$, $Y=\frac{E}{1-{\nu}^2}$. Moreover, the elastic parameters involve in the Young's modulus $E=193 ~{\rm GPa}$ and Poisson ratio $\nu=0.3$. In the simulation, the mass and elastic coefficient of each grain are as follows: $m=0.517~{\rm g}$ and $k=1.58 \times 10^{11}~{\rm N/m^{3/2}}$. The Verlet-velocity algorithm is adopted to update the position and velocity of grains. The simulation time step is that $dt=1.0 \times 10^{-9}~{\rm s}$. The mark of "+" represents the case of incident wave coming from the direction of BC, and "-" for that from the direction of MC. The letters of $I, T$, and $R$ are simplified forms of the incidence, transmission and reflection, respectively.

To trigger the SW, the first edge grain in the chain was given a certain impact velocity as a striker and the others stayed still  \cite{Nest2007APL90P261902, Sen2005PRL94P178002}. In our simulations, the striker velocity was set at the longitudinal direction of each chain. After changing the magnitude of striker velocity $v_{\rm imp}$ ranging from $0.01$ to $10.0~{{\rm m/s}}$, the same phenomena were reproduced. The width of generated SW was $11d$ and its amplitude was $v_{\rm max}/v_{\rm imp}=0.682$. In the following, the striker velocity was fixed at $v_{\rm imp}=1.0~{\rm m/s}$. Timing started up when the amplitude of SW just arrived at the grain 30. At this moment, the grain 30 got to the first maximum velocity and the velocities of grain 35 and grain 25 were zero.

\begin{figure}[htbp]\centering
\includegraphics[width=10.0 cm]{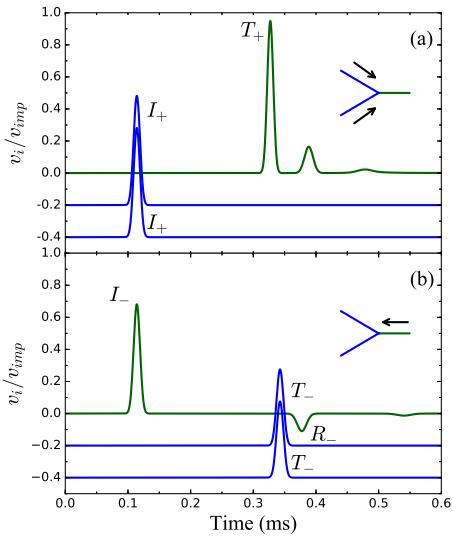}
\caption {(Color online) Temporal evolution of the velocities of grain TBC 15, BBC 15 and MC 15 at the branch angle of $\alpha=30^{0}$. The incident waves entered from the directions of (a) BCs and (b) MC, respectively. And the curves of TBC 15 and BBC 15 were offset downward by 0.2 and 0.4 for clarity, respectively.}
\label{fig:FigAn30}
\end{figure}

Firstly, we simulated the wave rectification of two SWs coming from the directions of TBC and BBC at the branch angle of $\alpha=30^{0}$, respectively. In Fig. \ref{fig:FigAn30}(a), the relationship between the velocity of grain TBC 15, BBC 15, and MC 15 and the time is given, respectively. Two incident SWs identically passed through the TBC 15 and BBC 15 and then went through the Y-shaped junction. The superposed wave entered into the MC in which a leading transmitted SW with higher amplitude than that of the incident SW was born and at least two smaller SWs were generated in order. In the BCs, the reflected wave was almost prohibited, and the data showed that there were very small negative velocities in the TBC and BBC. In Fig. \ref{fig:FigAn30}(b), the SW came from the direction of MC. Compared with the previous case, the system of YGC had a rectification effect on the SW. Both reflected and transmitted waves were obtained as shown in Fig. \ref{fig:FigAn30}(b). Due to the symmetry of system, the transmitted SWs in TBC and BBC are completely equal. More importantly, a larger fraction of incident energy shown in Fig. \ref{fig:FigAn30}(b) is reflected than that in Fig. \ref{fig:FigAn30}(a).

\begin{figure}[htbp]
\centering
\includegraphics[width=10.0 cm]{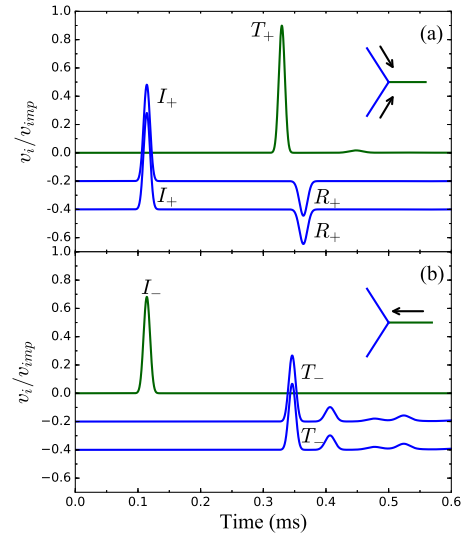}
\caption {(Color online) Temporal evolution of the velocities of grain TBC 15, BBC 15 and MC 15 at the branch angle of $\alpha=60^{0}$. The incident waves came from the directions of (a) BCs and (b) MC, respectively. And the data of TBC 15 and BBC 15 were offset downward by 0.2 and 0.4 for clarity, respectively.}
\label{fig:FigAn60}
\end{figure}

The same simulations were carried out on the YGC at the branch angle of $\alpha=60^{0}$ as shown in Fig. \ref{fig:FigAn60}. While the system shows the completely opposite properties of wave rectification compared to that in Fig. \ref{fig:FigAn30}. When the SW enters from the direction of BC, both reflected SW in the BC and transmitted SW in the MC are clearly observed shown in Fig. \ref{fig:FigAn60}(a). Similarly, the results shown in Fig. \ref{fig:FigAn60}(b) are almost the same to that shown in Fig. \ref{fig:FigAn30}(a) when the incident SW enters from the direction of BC. Several transmitted SWs are generated in the TBC and BBC, and while the reflected waves in the MC are not completely observed.

\begin{figure}[htbp]
\centering
\includegraphics[width=10.0 cm]{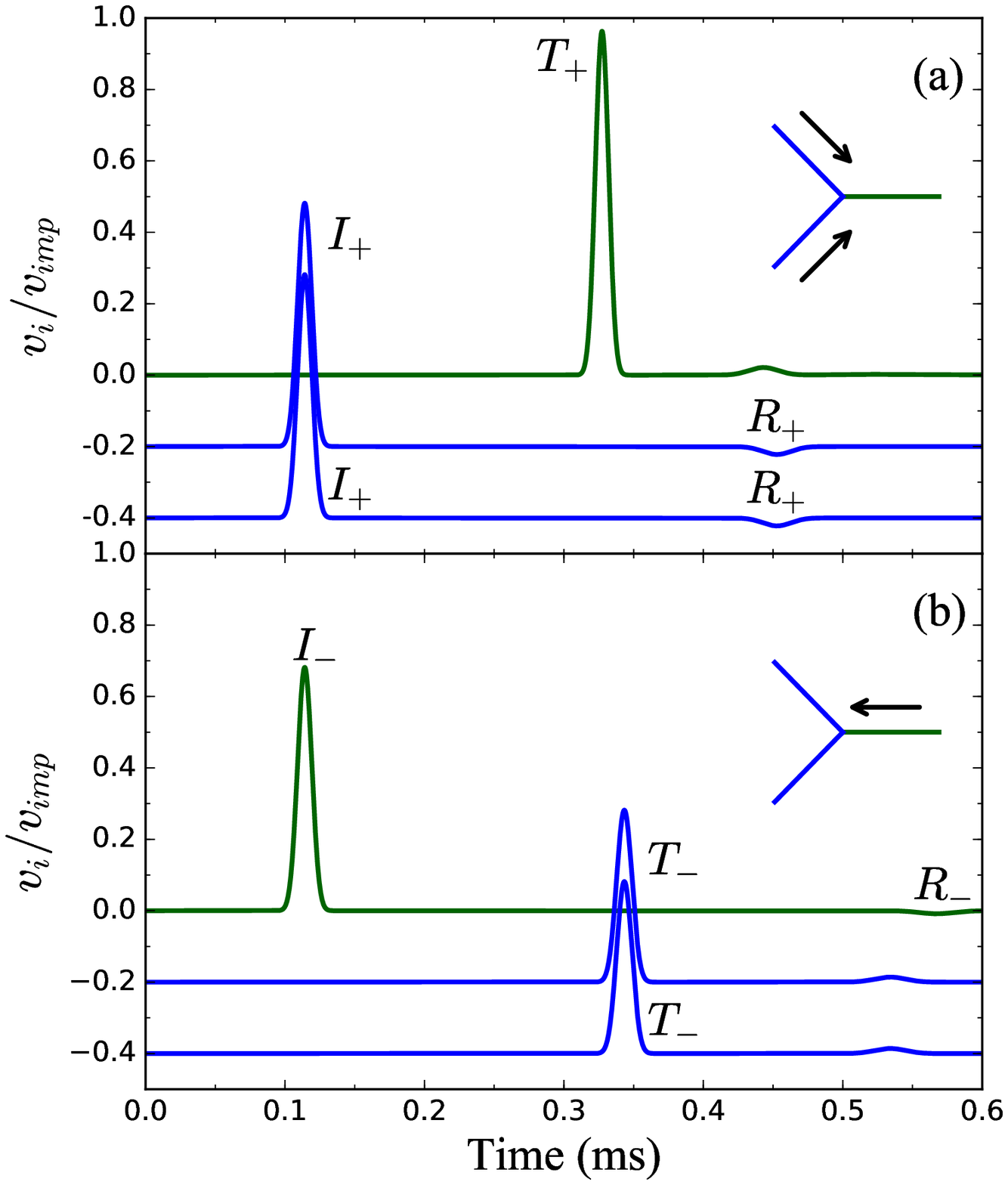}
\caption {(Color online) Temporal evolution of the velocities of grain TBC 15, BBC 15 and MC 15 at the branch angle of $\alpha=45^{0}$. The incident wave entered from the directions of (a) BCs and (b) MC, respectively. And the curves of TBC 15 and BBC 15were offset downward by 0.2 and 0.4 for clarity, respectively.}
\label{fig:FigAn45}
\end{figure}

The YGC with the branch angle of $\alpha=45^{0}$ was also simulated and the results were plotted in Fig. \ref{fig:FigAn45}. It is found that both transmitted and reflected waves generate when the incident waves come from the directions of BCs in Fig. \ref{fig:FigAn45}(a) and MC in Fig. \ref{fig:FigAn45}(b), respectively. Both cases show that the leading transmitted wave has higher amplitude than that of the reflected wave, which means that more incident energy is transmitted.

\begin{figure}[htbp]
\centering
\includegraphics[width=10.0 cm]{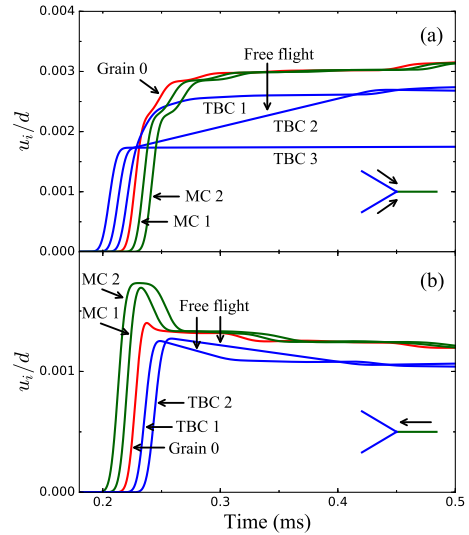}
\caption{(Color online) Temporal evolution of the displacement of grains around the Y-shaped junction. The same conditions as those in Fig. \ref{fig:FigAn30}(a)(b) were used in (a)and (b), respectively.}
\label{fig:FigUxTime30}
\end{figure}

The simulation results have shown that the SW rectification in the YGC can be controlled by adjusting the branch angle and incident direction. To understand the origin of wave rectification, Fig. \ref{fig:FigUxTime30} plots the temporal evolution of the displacement of grains around the Y-shaped junction under the same conditions as in Fig. \ref{fig:FigAn30}. The slope of displacement-time curve is the grain velocity. And the positive slope indicates that the grain keeps moving on, while the negative slope means that the grain is reflected back. In Fig. \ref{fig:FigUxTime30}, the interested is that the free flights appear at the straight lines of the displacement-time curves, such as, the curves of TBC 2 in Fig. \ref{fig:FigUxTime30}(a) and those of TBC 1 and TBC 2 in Fig. \ref{fig:FigUxTime30}(b). These free flights imply that there are gaps between the adjacent grains, see Movie 1 and Movie 2 in Supplemental Material \cite {Movie1&2}. Also the similar free flights and gaps are also observed under the conditions of branch angles of $\alpha=60^{0}$ and $\alpha=45^{0}$. The occurrence of the gap means that the SW breaks down and the grain-grain collision happens as reported in Refs. \cite{Nest2005PRL95P158702, Nest2013PRL111P048001}. Accompanying with the opening and closing of the gap, a series of transmitted and reflected waves generate at the same time.

The wave rectification might seem to result from the effects of both dispersion and collision. Taking the YGC with the branch angle of $\alpha=30^{0}$ for example, it might be regarded as a heavy-light chain by using the momentum conservation in horizontal direction \cite{Vergara2005PRL95P108002}. The case under the condition that two SWs are injected from the directions of TBC and BBC is considered firstly. As the superposed transmitted wave has larger amplitude and higher propagation velocity than those of the incident wave \cite{Nest1984JAMTP, Coste1997PRE56P6104}. And due to the different wave velocities in the BC and MC, the dispersion effect occurs and the incident wave and transmitted wave break down around the Y-shaped junction. In the meantime, a mass heterojunction is formed because of the mass mismatch as shown in Fig. \ref{fig:Figure1Model}(b). The action of incident wave seems to be a heavy ball \cite{Sen2007EPL77P24002, Sen2007GM10P13}. Thus the collision effect occurs where the velocity is decreased for the heavy ball in the BC, and the light ball in the MC gets a higher velocity. Both heavy ball and light ball keep moving forward after the collision. The light ball will collide with the next ball and its velocity is decreased. Then the light ball is hit by the heavy ball again and the secondary wave is generated. Therefore, the appearance of the gap and a train of secondary waves can be interpreted by the combined effects of dispersion and collision. When the mass of heavy grain is increased, both dispersion effect and collision effect take a positive role on the increasing of transmission. In reverse, when a SW enters from the direction of MC, the Y-shaped chain is analogue to a light-heavy chain. The incident wave is decomposed into two equal ones. The transmitted waves in the TBC and BBC have smaller amplitudes and lower propagation velocities. The breakdown of incident wave and transmitted wave occurs due to the dispersion effect. Similarly, the gap between grains appears and the collision effect leads to the reflection of incident light ball as it collides with a heavy ball. At the latter time, the reflected grain goes back to the chain and is reflected again. Then the continuous forward and backward collisions activate the secondary transmitted and reflected waves. Moreover, the mass decreasing of light ball can reduce the transmission.

Based on the simulations and analyses above, it is reasonable to simplify the SWs in the BCs and MC as three quasi-particles with the effective mass $m_{\rm eff}$ and effective velocity $v_{\rm eff}$ \cite{Daraio2012PRE85P036602, Sen2007GM10P13}. When the SW passes through the Y-shaped junction, both of the momentum in the horizontal direction and the kinetic energy are conserved.

\begin{align}
m_{\rm eff}v_{\rm eff} {\rm cos} {\alpha}+m_{\rm eff}v_{\rm eff}{\rm cos} {\alpha}= M_{\rm eff}V_{\rm eff},
\label{eq:MomCon}
\end{align}

\begin{align}
\frac {1}{2}m_{\rm eff}v_{\rm eff}^{2}+\frac {1}{2}m_{\rm eff}v_{\rm eff}^{2}= \frac {1}{2}M_{\rm eff}V_{\rm eff}^{2},
\label{eq:EnCon}
\end{align}
where $M_{\rm eff}$ and $V_{\rm eff}$ are the combined effective mass and effective velocity of BCs in the horizontal direction, respectively. For the case of the SW coming from BC, the mass factor can be defined as follows:

\begin{align}
\beta_{+}=\frac {M_{\rm eff}}{m_{\rm eff}}=2({\rm cos} \alpha)^{2},
\label{eq:MassFactor}
\end{align}

When the SW comes from the direction of MC, the mass factor is denoted as $\beta_{-}=1/ \beta_{+}$. By using the quasi-particle model, the mass mismatch at the Y-shaped junction results in an acoustic heterojunction. The YGC may be treated as three kinds of the heavy-light chain ($\beta_{+} > 1$, $\beta_{-} < 1$), light-heavy chain ($\beta_{+} < 1$, $\beta_{-} > 1$), and mono-dispersed chain ($\beta_{+} = \beta_{-} = 1$).

Followed the discussion of quasi-particle model, the transmission coefficient can be derived by the collision of two particles. When the incident SW comes from the direction of BC, the transmission coefficient can be obtained by:

\begin{align}
\gamma^{T}_{\pm}=\frac {E^{T}_{\pm}}{E^{I}_{\pm}}
=
\left\{\begin{array}{l}
~~~~1,~~~~~~ \beta_{\pm} \ge 1  \\
\frac {4\beta_{\pm}}{(1+\beta_{\pm})^{2}},~~\beta_{\pm} < 1  \\
\end{array},\right.
\label{eq:TransCoeff}
\end{align}%

where ${E^{I}_{\pm}}$ and ${E^{T}_{\pm}}$ are the incident energy and transmitted energy, respectively. For the case of the SW going through a heavy-light junction ($\beta_{\pm} \ge 1$), the heavy grain keeps moving forward after the collision, and all incident energy is transmitted through the Y-shaped junction and no energy is reflected. On the contrary, a fraction of incident energy is transmitted and the rest is reflected when the SW passes through a light-heavy junction ($\beta_{\pm} < 1$).

\begin{figure}[htbp]
\centering
\includegraphics[width=9.0cm]{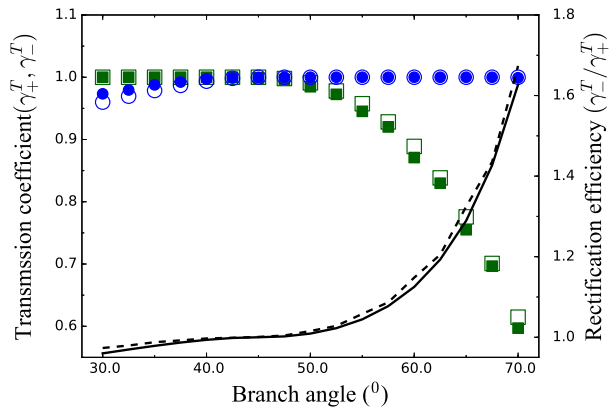}
\caption{(Color online) Relationships of the transmission coefficients and rectification efficiency with the branch angle. Square and circle symbols are for the results of incident SW coming from the directions of BC and MC, respectively. Solid and open symbols are for the transmission coefficients of the simulation and theoretical results, respectively. Solid and dashed lines are for the rectification efficiency of simulation and theoretical results, respectively.}
\label{fig:FigEnTRE}
\end{figure}

The propagation of SW can be efficiently controlled by the YGC with different branch angles. The relationship of the total transmission of energy flux with the branch angle is plotted in Fig. \ref{fig:FigEnTRE}. The simulation results are in good agreement with the theoretical predictions. As expected, the simulated YGC is identified as an effective GAD model. $\gamma^{T}_{+}$ and $\gamma^{T}_{-}$ are continuously modulated by adjusting the branch angle. When $\alpha\leq45^{0}$, the incident SW from the direction of BC has a higher transmission coefficient than that from the direction of MC, i.e., $\gamma^{T}_{+}>\gamma^{T}_{-}$. Conversely, when $\alpha>45^{0}$, the YGC allows more energy to pass through the Y-shaped junction as the SW is injected from the direction of MC, that is, $\gamma^{T}_{+}<\gamma^{T}_{-}$. The corresponding rectification efficiency is also plotted in Fig. \ref{fig:FigEnTRE},where the larger the branch angle is, the higher the rectification efficiency is.

In summary, we have numerically investigated the rectification effect on the solitary wave in a symmetric Y-shaped granular chain. The branch angle dependence of the transmission and reflection is identified by both simulation experiments and analytical predictions. The dynamics of solitary wave propagation in the Y-shaped granular chain is analogous to that of one dimensional granular chain with a heavy chain and a light chain. The combined effects of dispersion and collision result in the breakdown of incident and transmitted waves. And the increasing of branch angle is favorable for the increasing of rectification efficiency. The basic prototype of simulations for the Y-shaped granular chain offers the possibility of designing granular acoustic diodes, which might have potential application for the acoustic wave manipulation.

%\clearpage

This work is financially supported by the National Natural Science Foundation of China (Grant Nos. 11574153, 21574066, 11574150, 11334005, 11572178, 91634202) and Jiangsu Province Postdoctoral Science Foundation (Grant No. 1402007C).


\begin{thebibliography}{99}

\bibitem{Tongay2012PRL2P011002}
S. Tongay, M. Lemaitre, X. Miao, B. Gila, B. R. Appleton and A. F. Hebard, Phys. Rev. Lett.  {\bf 2}, 011002 (2012).

\bibitem{Zietek2016APL109P072406}
A. Zietek, P. Ogrodnik, W. Skowro\'{n}ski, F. Sobiecki, S. Van Dijken, and T. Stobiecki, Appl. Phys. Lett. {\bf 109}, 072406 (2012).

\bibitem{Lepri2003PR377P1}
S. Lepri, R. Livi, and A. Politi, Phys. Rep. {\bf 377}, 1 (2003).

\bibitem{Zhu2011PRL106P014301}
X. F. Zhu, B. Liang, W. W. Kan, X. X. Zou, and J. C. Cheng, Phys. Rev. Lett.  {\bf 106}, 014301 (2011).

\bibitem{Sen2011APL99P063501}
A. Breindel, D. Sun, and S. Sen, Appl. Phys. Lett.  {\bf 99}, 063510 (2011).

\bibitem{Sen2015APL107P244105}
M. A. Przedborski and S. Sen, Appl. Phys. Lett.  {\bf 107}, 244105 (2015).

\bibitem{Sen2008PR462P21}
S. Sen, J. Hong, J. Bang, E. Avalos and R. Doney, Phys. Rep. {\bf 462}, 21 (2008).

\bibitem{Vergara2005PRL95P108002}
L. Vergara, Phys. Rev. Lett. {\bf 95}, 108002 (2005).

\bibitem{Daraio2012PRE85P036602}
D. Ngo, F. Fraternali and C. Daraio, Phys. Rev. E {\bf 85}, 036602 (2012).

\bibitem{Daraio2014JMPS73P103}
A. Leonard, L. Ponson and C. Daraio, J. Mech. Phys. Solids {\bf 73}, 103 (2014).

\bibitem{HuangPRE2006}
D. C. Huang, G. Sun, and K. Q. Lu, Phys. Rev. E {\bf 74}, 061306 (2006).

\bibitem{HuangEPJE2013}
D. C. Huang, M. Lu, S. Sen, M. Sun, Y. D. Feng and A. N. Yang, Eur. Phys. J. E {\bf 36}, 41 (2013).

\bibitem{Kuwabara1987}
G. Kuwabara, and K. Kono, Japn. J. Appl. Phys. {\bf 26}, 1230 (1987).

\bibitem{Schafer1996}
J. Sch\"{a}fer, S. Dippel, and D. E Wolf, J. Phys. I {\bf 6}, 5 (1996).

\bibitem{Nest2007APL90P261902}
E.B. Herbold, and V. F. Nesterenko, Appl. Phys. Lett. {\bf 90}, 261902 (2007).

\bibitem{Sen2005PRL94P178002}
S. Job, F. Melo, A. Sokolow, and S. Sen, Phys. Rev. Lett. {\bf 94}, 178002 (2005).

\bibitem{Movie1&2}
See Supplemental Material for additional movies. In the movies, the displacements of grains around the Y-shaped junction are amplified 50 times for clarity.

\bibitem{Nest2005PRL95P158702}
V. F. Nesterenko, C. Daraio, E. B. Herbold, and S. Jin, Phys. Rev. Lett. {\bf 111}, 048001 (2013).

\bibitem{Nest2013PRL111P048001}
A. M. Tichler, L. R. G\'{o}mez, N. Upadhyaya, X. Campman, V. F. Nesterenko, and V. Vitelli, Phys. Rev. Lett. {\bf 111}, 048001 (2013).

\bibitem{Nest1984JAMTP}
V. F. Nesterenko, J. Appl. Mech. Phys. {\bf 24} 734 (1984).

\bibitem{Coste1997PRE56P6104}
C. Coste, E. Falcon, and S. fauve, Phys. Rev. E {\bf 56}, 6104 (1997).

\bibitem{Sen2007EPL77P24002}
A. Sokolow, E. G. Bittle, and S. Sen, Europhys. Lett. {\bf 77}, 24002 (2007).

\bibitem{Sen2007GM10P13}
S. Job, F. Melo, A. Sokolow, and S. Sen, Granul. Matt. {\bf 10}, 13 (2007).

\end{thebibliography}
\end{document}